\documentclass[conference,letterpaper]{IEEEtran}
\usepackage[utf8]{inputenc} 
\usepackage[T1]{fontenc}
\usepackage{url}
\usepackage{ifthen}
\usepackage{cite}
\usepackage[cmex10]{amsmath} 

\usepackage{amsfonts}
\usepackage{amssymb}
\usepackage{amsthm}
\usepackage{algorithmicx}
\usepackage{algpseudocode}
\usepackage{algorithm}
\usepackage{subcaption}
\usepackage{graphicx}
\usepackage{cite}
\usepackage{calc}
\usepackage{color}
\usepackage{epsfig}
\usepackage{setspace}
\usepackage{multirow}
\usepackage{mathtools, cuted}
\usepackage{epstopdf}
\usepackage{lipsum}
\usepackage{mathtools}
\usepackage{cuted}
\usepackage{caption}

\newtheorem{defn}{Definition}
\newtheorem{thm}{{\cal T}heorem}
\newtheorem{cor}[thm]{Corollary}
\newtheorem{prop}{Proposition}
\newtheorem{lem}[thm]{Lemma}
\newtheorem{conj}[thm]{Conjecture}
\newtheorem{constr}[thm]{Construction}
\newtheorem{note}{Remark}
\newcommand{\bit}{\begin{itemize}}
	\newcommand{\eit}{\end{itemize}}
\newcommand{\bcor}{\begin{cor}}
	\newcommand{\ecor}{\end{cor}}
\newcommand{\beq}{\begin{equation}}
\newcommand{\eeq}{\end{equation}}
\newcommand{\beqn}{\begin{equation}}
\newcommand{\eeqn}{\end{equation}}
\newcommand{\bea}{\begin{eqnarray}}
\newcommand{\eea}{\end{eqnarray}}
\newcommand{\bean}{\begin{eqnarray*}}
	\newcommand{\eean}{\end{eqnarray*}}
\newcommand{\ben}{\begin{enumerate}}
	\newcommand{\een}{\end{enumerate}}
\newcommand{\bdefn}{\begin{defn}}
	\newcommand{\edefn}{\end{defn}}
\newcommand{\bnote}{\begin{note}}
	\newcommand{\enote}{\end{note}}
\newcommand{\bprop}{\begin{prop}}
	\newcommand{\eprop}{\end{prop}}
\newcommand{\blem}{\begin{lem}}
	\newcommand{\elem}{\end{lem}}
\newcommand{\bthm}{\begin{thm}}
	\newcommand{\ethm}{\end{thm}}
\newcommand{\bconj}{\begin{conj}}
	\newcommand{\econj}{\end{conj}}
\newcommand{\bconstr}{\begin{constr}}
	\newcommand{\econstr}{\end{constr}}
\newcommand{\bpf}{\begin{proof}}
	\newcommand{\epf}{\end{proof}}
\newcommand{\bprf}{{\em Proof: }}
\newcommand{\eprf}{\hfill $\Box$}

\newcommand{\calc}{\mbox{$\mathcal{C}$}}

\newcommand{\params}{\mbox{$(a,b,\tau)$}}

\IEEEoverridecommandlockouts 
\begin{document}
	
	\title{Generalized Simple Streaming Codes \\ from MDS Codes} 
\author{%
	\IEEEauthorblockN{Vinayak Ramkumar, Myna Vajha, P. Vijay Kumar}
	\IEEEauthorblockA{
		Department of Electrical Communication Engineering, IISc Bangalore \\ \{vinram93, mynaramana, pvk1729\}@gmail.com}
	
\thanks{This  research  is  supported by  the J C Bose National Fellowship JCB/2017/000017.}
}

	\maketitle

\begin{abstract}
	
Streaming codes represent a packet-level FEC scheme for achieving reliable, low-latency communication. In the literature on streaming codes, the commonly-assumed Gilbert-Elliott channel model, is replaced  by a more tractable, delay-constrained, sliding-window (DCSW) channel model that can introduce either random or burst erasures. The known streaming codes that are rate optimal over the DCSW channel model are constructed by diagonally embedding a scalar block code across successive packets. These code constructions have field size that is quadratic in the delay parameter $\tau$ and have a somewhat complex structure with an involved decoding procedure. This led to the introduction of simple streaming (SS) codes in which diagonal embedding is replaced by staggered-diagonal embedding (SDE). The SDE approach reduces the impact of a burst of erasures and makes it possible to construct near-rate-optimal streaming codes using Maximum Distance Separable (MDS) code having linear field size. The present paper takes this development one step further, by retaining the staggered-diagonal feature, but permitting the placement of more than one code symbol from a given scalar codeword within each packet. These generalized, simple streaming codes allow us to improve upon the rate of SS codes, while retaining the simplicity of working with MDS codes. We characterize the maximum code rate of streaming codes under a constraint on the number of contiguous packets over which symbols of the underlying scalar code are dispersed.   Such a constraint leads to simplified code construction and reduced-complexity decoding. 
	\end{abstract}
\section{Introduction}
The availability of a reliable, low-latency communication system is key to many envisaged 5G applications such as telesurgery,  augmented and virtual reality.  Packet drops are commonplace in a communication network and can arise due to congestion in the network, a weak wireless link or delayed packet arrival. There is need for a communication scheme that can recover from such packet losses in a timely fashion \cite{5GAmericas}. 
Packet duplication amounts to using a repetition code and is clearly inefficient. Instantaneous-feedback-based approaches such as ARQ incur an undesired round-trip delay. Streaming codes represent a packet-level Forward Error Correction (FEC) scheme that is both efficient and of low latency. 

The study of streaming codes began in \cite{MartSunTIT04,MartTrotISIT07} where packet-level FEC codes capable of handling an erasure burst within a decoding-delay-window were investigated. Here, encoding is carried out using a packet-expansion framework, a feature that has been retained in the subsequent literature as well, including the present paper. At any time $t$, if $\underline{m}(t) \in \mathbb{F}_q^k$ is the message packet, then the coded packet $\underline{x}(t) = \left[\underline{m}(t)^T \ \underline{p}(t)^T\right]^T \in \mathbb{F}_q^n$, where $\underline{p}(t)\in \mathbb{F}_q^{n-k}$ represents parity.  A decoding-delay constraint of $\tau$, is construed as requiring that $\underline{m}(t)$ be decoded by time $(t+\tau)$.
 
In  \cite{BadrPatilKhistiTIT17}, a delay-constrained sliding-window (DCSW) channel model was introduced as a tractable deterministic approximation to the more realistic Gilbert-Elliott erasure channel model \cite{gilbert,elliott,HasHoh,VajRamJhaKum} that is capable of causing burst and random erasures. An $(a,b,w,\tau)$ DCSW channel imposes a decoding-delay constraint of $\tau$ and permits either at most $a$ random erasures or else, a burst of $b$ erasures within any sliding window of size $w$ time slots, where $0<a \leq b \leq \tau$. Without loss of generality we can set $w=\tau+1$ (see \cite{BadrPatilKhistiTIT17,NikDeepPVK}). A  packet-level code will be referred to as an \params\ streaming code if it can recover under decoding-delay $\tau$ from all the admissible erasure patterns of the $(a,b,w=\tau+1,\tau)$ DCSW channel. An upper bound on the rate of \params\ streaming code was derived in \cite{BadrPatilKhistiTIT17} and codes achieving this rate for all possible parameters were first presented in \cite{FongKhistiTIT19,NikPVK}, thereby characterizing the optimal rate $R_{\text{opt}}(a,b,\tau) = \frac{\tau+1-a}{\tau+1-a+b}$. These initial rate-optimal codes are over a finite field of size exponential in $\tau$. Currently, the known \cite{NikDeepPVK,KhistiExplicitCode} rate-optimal streaming code constructions have field size $q$ that is quadratic in the delay parameter $\tau$. In \cite{DudFongKhi}, rate-optimal streaming codes with rate at least $\frac{1}{2}$ are constructed by combining Maximum Distance Separable (MDS) codes and Maximum Rank Distance codes, but this construction requires a large field size of $O((2a)^{\tau})$. Streaming codes for variable size message packets are studied in \cite{RudRas}. Other FEC schemes suitable for streaming setting can be found in \cite{AdlCas,LeongHo,LeoQurHo,MalMedYeh,bats,Shokrollahi,IyeSUW,FelZig,DraKhi}.

Constructions of streaming codes were originally based on diagonal embedding (DE) of a scalar block code \calc\ within the packet stream and these codes have a complex structure that could make implementation challenging. Under DE, every diagonal in the packet stream is a codeword in \calc. In \cite{NikRamVajKum}, a variant of DE called staggered-diagonal embedding (SDE) is introduced which reduces the impact of burst erasures. 
\begin{figure*}
	\centering
	\begin{subfigure}[b]{0.31\textwidth}
		\centering
		\includegraphics[width=0.75\textwidth]{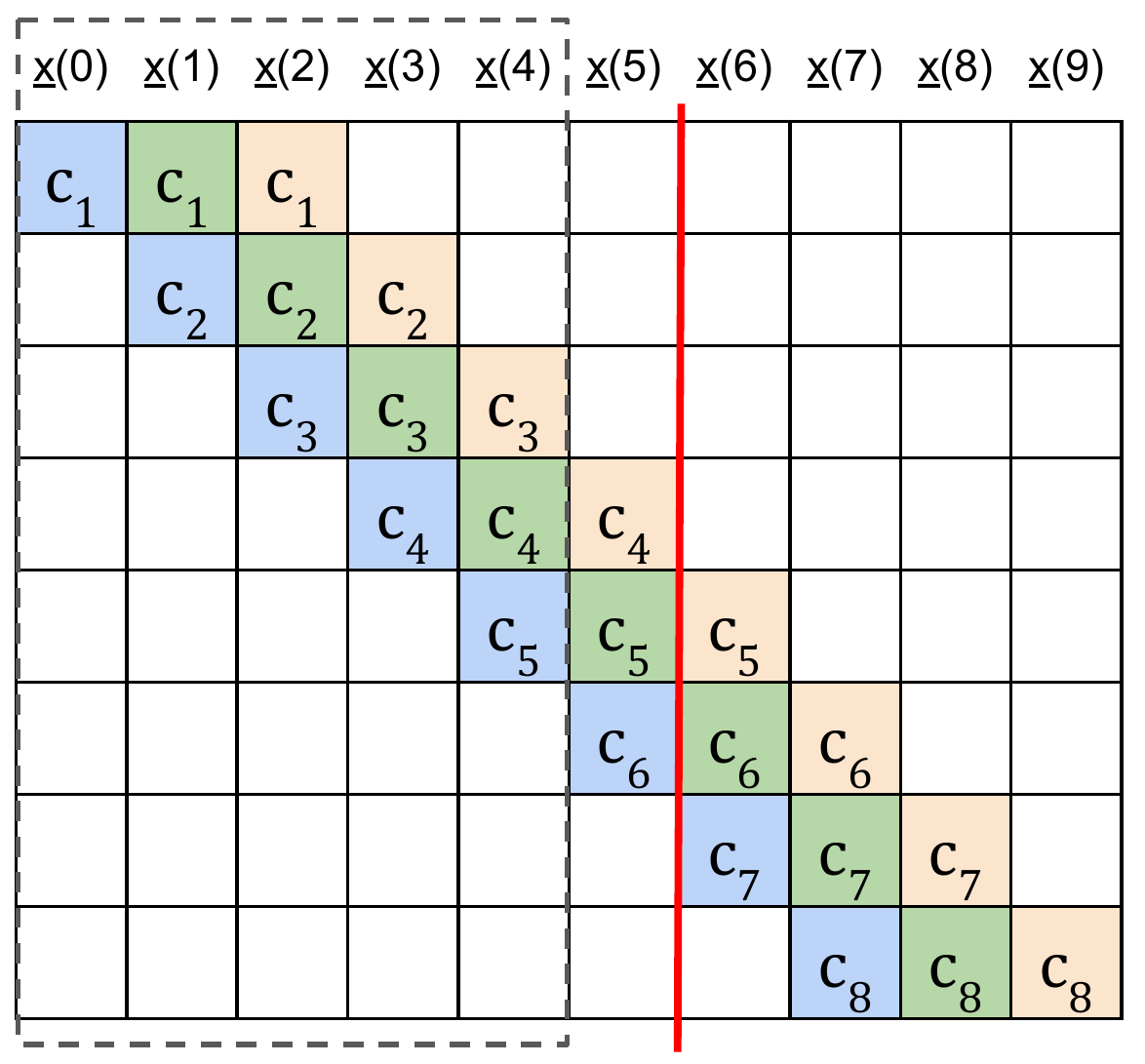}   
		\caption{DE associated of an $[8,3]$ scalar code over $\mathbb{F}_q$, $q\geq 25$, leading to a rate-optimal streaming code~\cite{NikDeepPVK}. }
		\label{Fig:DE}
 \ \\
	\end{subfigure}
	\hfill
	\begin{subfigure}[b]{0.31\textwidth}
		\centering
		\includegraphics[width=0.75\textwidth]{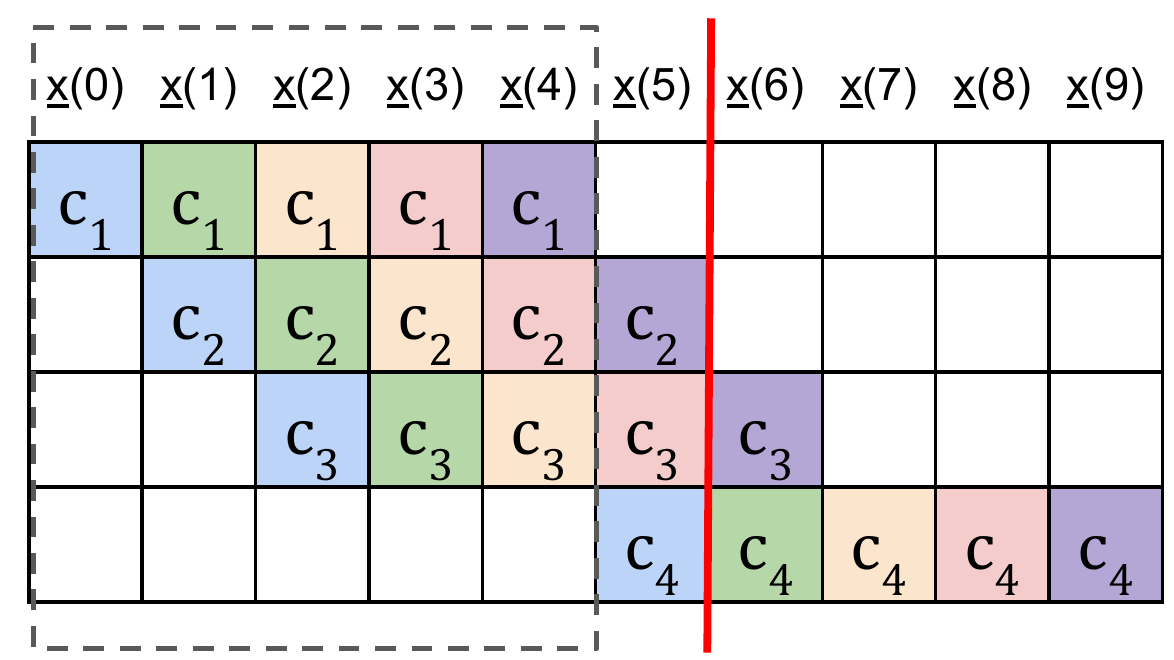}   
		\caption{SDE of a $[4,1]$ scalar MDS code leading to a SS code~\cite{NikRamVajKum} having rate $\frac{1}{4}$.}
		\label{Fig:SDE}
		\ \\ \ \\ \ \\  \
	\end{subfigure}
	\hfill
	\begin{subfigure}[b]{0.31\textwidth}
		\centering
		\includegraphics[width=0.75\textwidth]{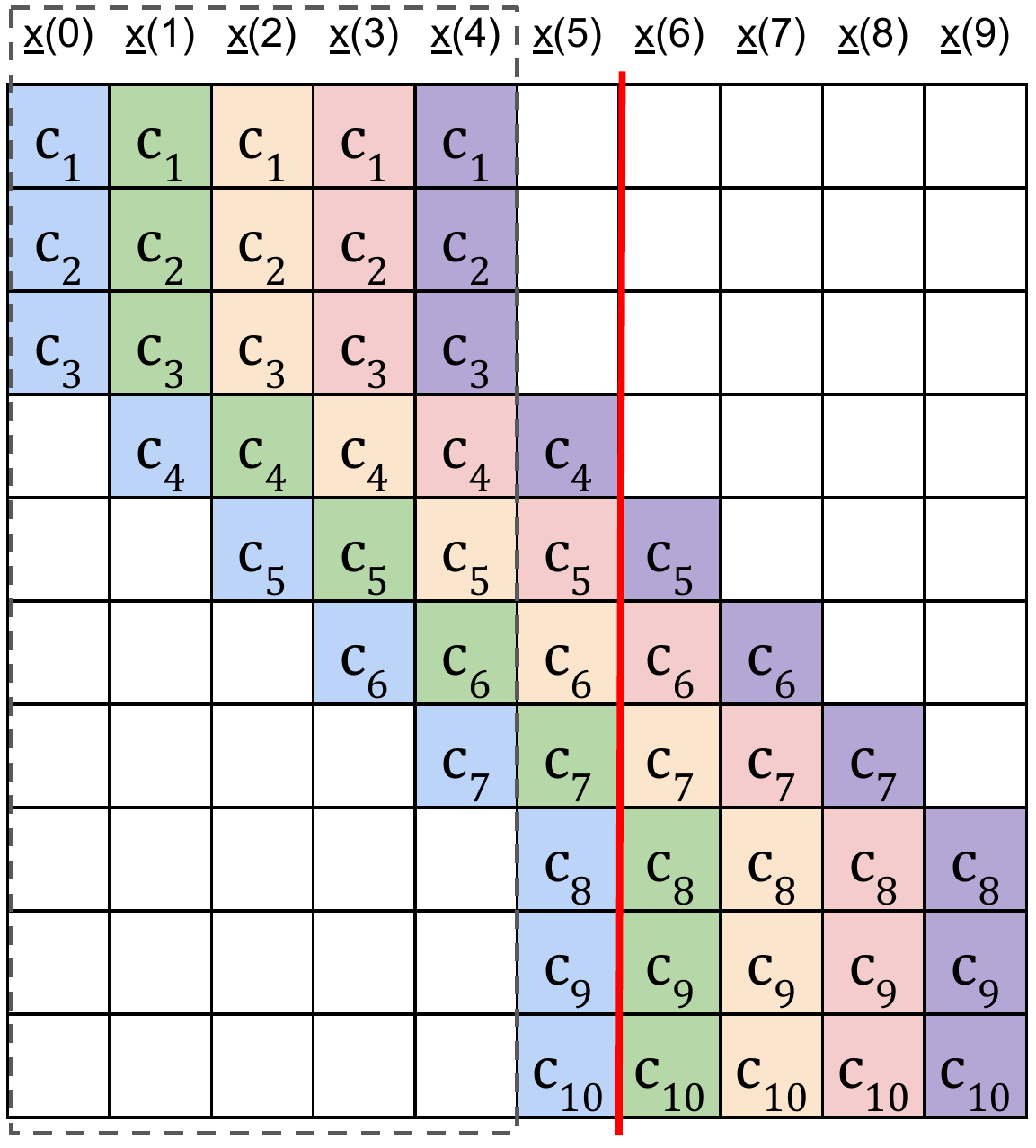}
		\caption{GSDE of a $[10,3]$ scalar MDS code with dispersion vector $(3,1,1,1,1,3)$ leading to a GSS code of rate $\frac{3}{10}$.} 
		\label{Fig:GSDE}
	\end{subfigure}
	\vspace{-0.05in}
	\caption{Three constructions of $(a=3,b=5,\tau=5)$ streaming code. Each column represents a coded packet and symbols bearing the same color belong to the same scalar codeword. The black dotted box indicates a burst erasure of $b=5$ packets and red line is the decoding deadline for packet $0$.}
	\vspace{-0.25in}
\end{figure*}
 In SDE framework a coded packet can contain atmost one symbol from a codeword. Dispersion span $N$ is defined as the number of contiguous packets over which code symbols of the underlying scalar code \calc\ are dispersed. If $N \leq \tau+1$, then decoding is simplified as each codeword in \calc\ can be block decoded. For example, in Fig.~\ref{Fig:SDE} the dispersion span is $N=\tau+1=6$. Simple Streaming (SS) codes are constructed by SDE with $N \le \tau+1$ and the underlying scalar code \calc\ is chosen to be a Reed-Solomon or other MDS code over a finite field $\mathbb{F}_q$ of size $q=O(\tau)$. MDS codes are widely used in practice and low-complexity algorithms and libraries for encoding and decoding MDS codes are available~\cite{PlaGre,isal}. The maximum possible rate of a streaming code obtained through SDE with $N \leq \tau+1$ is characterized in \cite{NikRamVajKum} and SS codes achieve it. In \cite{RamVajNikPVK}, SDE with dispersion span larger than $\tau+1$ is explored. 


In the present paper, we take the development of streaming codes one step further by allowing within the SDE framework, the embedding of more than one code symbol form the underlying scalar code \calc\ within a coded packet.  We refer to this form of embedding as Generalized SDE (GSDE). Under the condition that the dispersion span $N$ satisfies $N \leq \tau+1$, we characterize the maximum possible rate of an streaming code constructed via GSDE. We show that the underlying scalar code \calc\ can always be chosen to be an MDS code and hence these codes are simpler to implement. The resultant streaming codes are referred to as Generalized Simple Streaming (GSS) codes. Clearly, the rate of a GSS code can be no smaller than that of a SS code and we 
characterize the improvement in code rate over than of an SS code.
	\vspace{-0.05in}
\subsection{Motivating Example: $(a=3, b=5, \tau=5)$} 
For  $(a=3, b=5, \tau=5)$, the best possible rate of a streaming code is $R_{opt}= \frac{3}{8}$. This rate can be achieved by DE of $[8,3]$ scalar code presented in \cite{NikDeepPVK}, see Fig.~\ref{Fig:DE}. A field of size $\ge 25$ is required and this code is not easy to implement. The rate achievable using an SS code is $R_{SS} = \frac{1}{4}$, obtained by SDE of $[4,1]$ MDS code as illustrated in  Fig.~\ref{Fig:SDE}. We now describe the construction of an $(a=3, b=5, \tau=5)$ streaming code constructed using GSDE of an $[n=10,k=3]$ MDS code $\mathcal{C}$ having rate $\frac{3}{10}$, as shown in Fig.~\ref{Fig:GSDE}. Here the $n=10$ code symbols are dispersed across $N=6$ consecutive packets. The embedding can be described using dispersion vector $(3,1,1,1,1,3)$ which indicates the number of symbols assigned to successive packets within which the symbols of the MDS codeword are embedded. If $\underline{x}(t)=[x_1(t) \dots x_{10}(t)]^T$ denotes the coded packet at time $t$, then {\footnotesize $\big{(}x_1(t),x_2(t),x_3(t),x_4(t+1),x_5(t+2),x_6(t+3),x_7(t+4), x_8(t+5),x_9(t+5),x_{10}(t+5) \big{)}$} is a codeword of $\mathcal{C}$, for all $t$.   
It can be seen that the erasure of any $a=3$ packets or $b=5$ consecutive packets will result in the loss of at most $(n-k) = 7$ code symbols which is within the erasure-recovery capability of the scalar MDS code. Since $N=\tau+1$ here, the decoding-delay constraint is trivially satisfied. 
\begin{figure}
	\centering
	\begin{subfigure}[b]{0.24\textwidth}
		\centering
		\includegraphics[width=1.05\textwidth]{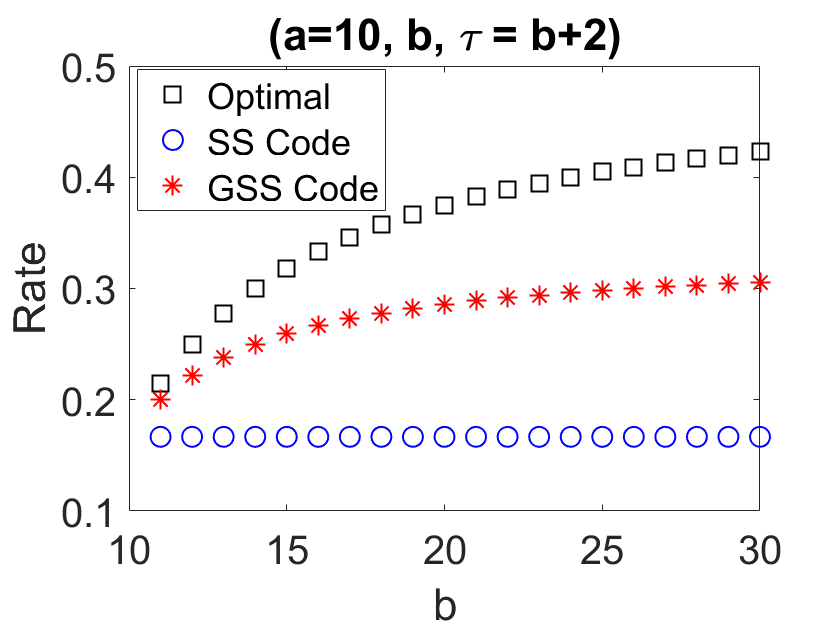}   
	\end{subfigure}
	\begin{subfigure}[b]{0.24\textwidth}
		\centering
		\includegraphics[width=1.05\textwidth]{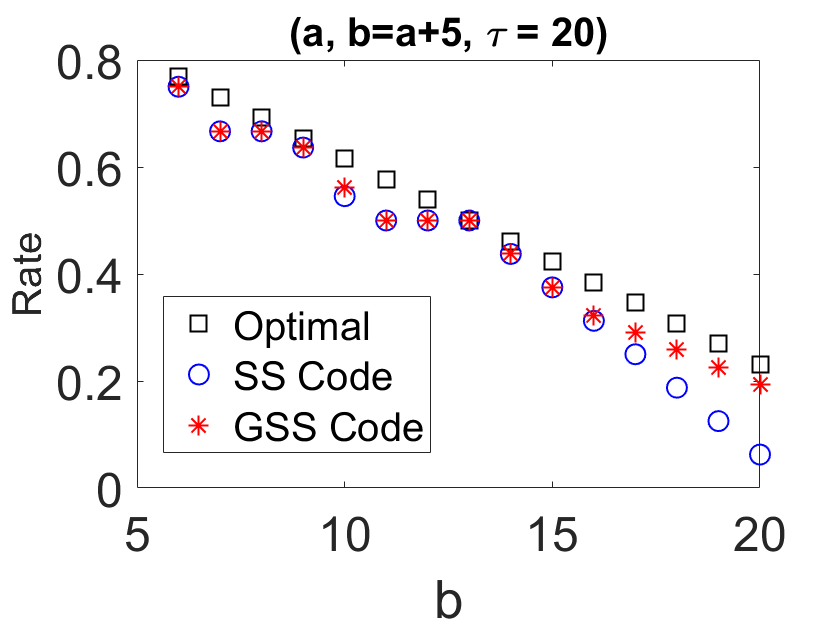}
	\end{subfigure}
	\caption{Rate comparison between rate-optimal streaming codes in \cite{NikPVK,FongKhistiTIT19,NikDeepPVK,KhistiExplicitCode}, SS codes in \cite{NikRamVajKum} and GSS codes presented here.}
	\label{Fig:rate}
	\vspace{-0.25in}
\end{figure}
This example illustrates the possibility of improving upon the rate of  SS codes by embedding more than one symbol of an MDS codeword within a single packet. Fig.~\ref{Fig:rate} shows the rate of  various streaming code constructions for two different parameter settings and the rate gain of GSS code over SS code can be observed in it.

The GSDE framework is formally introduced in Section~\ref{Sec:GSDE} and the construction of a GSS code is linked to an interesting combinatorial problem.  An upper bound on the rate of streaming codes constructed through GSDE with $N \le \tau+1$ is derived in Section~\ref{Sec:bound}. The streaming code constructions are presented in Section~\ref{Sec:constr} and it is shown that the rate bound derived in Section~\ref{Sec:bound} is always achievable.
	\vspace{-0.05in}
\section{Generalized Staggered Diagonal Embedding} \label{Sec:GSDE}
Let $\mathcal{C}$ be an $[n,k]$ scalar linear code which is systematic, with message symbols appearing as first $k$ code symbols. Let $\underline{\Delta}=(n_1,\dots,n_N)$ be an $N-$tuple of non-negative integers with $\sum\limits_{i=1}^{N}n_i=n$. With out loss of generality we assume $n_1 \ne 0$. We will refer to $\mathcal{C}$ as the base code and $\underline{\Delta}$ as the dispersion vector. The parameter $N$ is referred to as the dispersion span. Let the coded packet at time $t$ be denoted by $\underline{x}(t)=[x_1(t),\dots, x_n(t)]^T$. Set $m_j=\sum\limits_{i=1}^{j}n_i$ for all $j \in [1,N]$.  We say that the packet-level code is constructed by GSDE of base code $\mathcal{C}$ with dispersion vector $\underline{\Delta}$ if for every $t$,  
\bean
\resizebox{.48 \textwidth}{!} 
{ 
$\Big( \big(x_{i}(t+j-1), i \in [m_{j-1}+1,~m_j] \big), j \in [1,N] \Big)$
} 
\eean
are code symbols of a codeword in $\mathcal{C}$.  
The packet-level code thus obtained has rate $\frac{k}{n}$. Under the GSDE framework, symbols from a codeword in the base code are dispersed across at most $N$ successive  packets. Consider a codeword $\underline{c} \in \mathcal{C}$ and let $[t+1,t+2,\dots,t+N]$ be the time indices of the $N$ consecutive packets across which this codeword is dispersed. Then, coded packet $(t+i)$ contains $n_{i}$ code symbols of $\underline{c}$. The GSDE framework can be viewed as a generalization of DE and SDE frameworks. For all $i \in [1,N]$, $n_i=1$ for DE, whereas $n_i$ can take values only in $\{0,1\}$ for SDE. In GSDE, we allow $n_i$ to be any non-negative integer.  

In this paper we focus on the $N \le \tau+1$ case. This restriction ensures that it suffices to decode each scalar codeword in the conventional block-decoding way, i.e., decode all underlying message symbols after receiving all code symbols belonging to unerased packets. We note that since $n_i=0$ is permitted for $i>1$, $N \le \tau+1$ is equivalent to setting $N=\tau+1$. Throughout the remainder of the paper, when we speak of GSDE, we mean GSDE with $N=\tau+1$. We will now reduce the problem of streaming code construction using GSDE with $N=\tau+1$ to a combinatorial problem.
\bdefn
Given positive integers $a,b,\tau, n$ and  $r$, we say that 
$\underline{\Delta} = (n_1,\dots,n_{\tau+1})$, $n_i \ge 0$,  is an $(a,b,\tau,n, r)-$dispersion vector if:
\bea \label{Eq:total_sum}
\sum\limits_{i=1}^{\tau+1} n_i = n,
\eea 
\bea
 \label{Eq:A} 
\sum\limits_{i \in A} n_i \le r,~~~\forall A \subseteq [1,\tau+1] ~\text{with}~ |A| =a,\\
\label{Eq:B}
 \sum\limits_{i=j}^{j+b-1} n_i \le r, ~~~\forall j \in [1,\tau+2-b],
\eea
and additionally at least one inequality in \eqref{Eq:A} or \eqref{Eq:B} holds with equality. 
We attach a rate $R$ to an $(a,b,\tau,n,r)$-dispersion vector, given by $R=\frac{n-r}{n}$.
\edefn
For example, $(3,1,1,1,1,3)$ is a $(3,5,5,10,7)-$dispersion vector. Now let $\underline{\Delta}$ be an $(a,b,\tau,n, r)$-dispersion vector. Pick an $[n,k=n-r]$ MDS code $\mathcal{C}$ and construct a packet-level code by GSDE of $\mathcal{C}$ with dispersion vector $\underline{\Delta}$. We now argue that the packet-level code thus constructed is an $(a,b,\tau)$ streaming code. An $[n,n-r]$ MDS code can recover from erasure of any $r$ code symbols. If any arbitrary $a$ coded packets are erased, then by  \eqref{Eq:A} at most $r$ symbols of any codeword of $\mathcal{C}$ are erased. If $b$ consecutive packets are erased, then once again, no more than $r$ code symbols are lost from any codeword, since  \eqref{Eq:B} holds for $\underline{\Delta}$. Thus \eqref{Eq:A} guarantees recovery from $a$ random erasures whereas \eqref{Eq:B} assures recovery in the presence of a burst of $b$ erasures. The delay constraint is trivially met since symbols of a codeword are spread across consecutive $\tau+1$ packets. Hence, we have the following Lemma.   
\blem \label{Lem:GSDE1}  If there exists an $(a,b,\tau,n, r)$-dispersion vector $\underline{\Delta}$, then an $(a,b,\tau)$ streaming code of rate $R=\frac{n-r}{n}$ can be constructed through GSDE of $[n, n-r]$ MDS code with dispersion vector $\underline{\Delta}$.   
\elem
We complete the description of relation between dispersion vector and streaming code construction through GSDE by establishing the converse. 
\blem \label{Lem:GSDE2}  If GSDE  of an $[n,n-r]$ base code $\calc$ with dispersion vector  $\underline{\Delta}$ results in an $(a,b,\tau)$ streaming code, then $\underline{\Delta}$ is an $(a,b,\tau,n,s)$-dispersion vector, where $s \le r$.
\elem
\bprf
An $[n,n-r]$ code can not recover from more than $r$ erasures. Hence, from the $b$-burst erasure and $a$-random erasure correction property of an $(a,b,\tau)$ streaming code, it follows that $\underline{\Delta}$ is an $(a,b,\tau,n, \le r)$-dispersion vector. 
\eprf 

Let $R_{\max}(a,b,\tau)$ denote the maximum rate possible of an $(a,b,\tau)$ streaming codes constructed through GSDE with $N=\tau+1$. It can be inferred from Lemma \ref{Lem:GSDE2} that the rate of a GSDE based streaming code can not exceed the rate associated to the underlying dispersion vector. Thus, from Lemma \ref{Lem:GSDE1} and Lemma \ref{Lem:GSDE2} it follows that $R_{\max}(a,b,\tau)$ can be determined by finding out the maximum rate  associated to an $(a,b,\tau,n, r)$-dispersion vector, i.e.,
\bean 
\resizebox{.5 \textwidth}{!} 
{
$R_{\max}(a,b,\tau) =\max \left\{\frac{n-r}{n}  ~\Big{|}~  \exists ~\text{an}~  \{a,b,\tau,n,r\}-\text{dispersion vector} \right\}$.
} 
\eean 
In our search for maximum rate under GSDE, it suffices to restrict our attention to MDS codes.  This can be seen as follows.   Let $\mathcal{C}$ be an $[n,n-r]$ scalar code that is not an MDS code. Consider an $(a,b,\tau)$ streaming code of rate $\frac{n-r}{n}$ constructed by GSDE of $\mathcal{C}$ using dispersion vector $\underline{\Delta}$. By Lemma~\ref{Lem:GSDE2}, $\underline{\Delta}$ is necessarily an $(a,b,\tau,n, s)$-dispersion vector for some $s \le r$.  The GSDE of an $[n,n-s]$ MDS code with dispersion vector $\underline{\Delta}$ will result in an $(a,b,\tau)$ streaming code of rate $\frac{n-s}{n} \ge \frac{n-r}{n}$.  Thus it is not possible to outperform an MDS base code in terms of rate.

 We next state our main result, which is the characterization of the maximum possible rate of streaming codes constructed via GSDE and with $N=\tau+1$.
\bthm \label{Thm:main}
Let $\tau+1=mb+\delta$, where $m \in \mathbb{N}, \delta \in [0,b-1]$. Then, $R_{\max}(a,b,\tau)= \frac{m-1+\mu}{m+\mu}$ where
\bean 
\mu  =
\begin{cases} 
	\frac{b-a+m\delta}{(m+1)b-a} ~~~\text{if}~~~ a>(m+1)\delta>0,\\ \ \\
	\frac{\min\{\delta,a\}}{a} ~~~\text{else}.
\end{cases}
\eean  
\ethm
The rate upper bound, corresponding to the theorem above, is derived in Section~\ref{Sec:bound}. Constructions of dispersion vectors that achieve this rate are presented in Section~\ref{Sec:constr}. 
	\vspace{-0.05in}
\section{Rate Upper Bound} \label{Sec:bound}
We establish the rate upper bound separately for two regimes, in  Theorems~\ref{Thm:upper_bound1} and ~\ref{Thm:upper_bound2} respectively.
\bthm \label{Thm:upper_bound1}
Let $\tau+1= mb+ \delta$, where $m \in \mathbb{N}, \delta \in [0,b-1]$. If $a \le (m+1)\delta$ or $\delta=0$, then $R_{\max}(a,b,\tau) \le R_{SS}(a,b,\tau) = \frac{m-1+\frac{\min\{\delta,a\}}{a}}{m+\frac{\min\{\delta,a\}}{a}}$.
\ethm
\bprf 
Let $\underline{\Delta}=(n_1,\dots,n_{\tau+1})$ be an $(a,b,\tau,n, r)$-dispersion vector  with rate $R_{\max}(a,b,\tau)$.  
Suppose $\delta \ge a$. Since $\underline{\Delta}$ obeys \eqref{Eq:B} we have
\bean 
\sum\limits_{i=jb+1}^{(j+1)b}n_i \le r,\forall j \in [0,m-1]
~~\text{and}~~ \sum\limits_{i=mb+1}^{mb+\delta}n_i \le r.
\eean  
Hence $n=\sum\limits_{i=1}^{\tau+1}n_i \le (m+1)r$
and therefore  $R_{\max}(a,b,\tau)=\frac{n-r}{n} \le \frac{m}{m+1} = R_{SS}(a,b,\tau)$ when $\delta \ge a$. Proof for remaining parameter ranges can be found in Appendix~\ref{proof_upper_bound}.  
\eprf 

To derive rate upper bound for the case $a>(m+1)\delta>0$, we first show the existence of a maximum-rate dispersion vector having a specific structure.   
\blem \label{Lem: sum_r} For any valid $\{a,b,\tau\}$, there exists an $(a,b,\tau,n, r)$-dispersion vector  $\underline{\Delta}=(n_1,\dots,n_{\tau+1})$ of rate $R_{\max}(a,b,\tau)$ with $\sum\limits_{i=t_0}^{t_0+b-1} n_i = r$ for some $t_0 \in [1,\tau+2-b]$.
\elem  
 
\bprf 
See Appendix~\ref{proof_sum_r}.
\eprf

From Lemma~\ref{Lem: sum_r}, it can be inferred that in order to determine $R_{\max}(a,b,\tau)$ one needs to look at only the $(a,b,\tau,n,r)$-dispersion vectors for which \eqref{Eq:B} is satisfied with equality for at least one case.
\bthm \label{Thm:upper_bound2}
Let $\tau+1= mb+ \delta$, where $m \in \mathbb{N},\delta \in [0,b-1]$. If $\delta \neq 0$ and $a > (m+1)\delta$, then $R_{\max}(a,b,\tau) \le \frac{m-1+\theta}{m+\theta}$, where $\theta= \frac{b-a+m\delta}{(m+1)b-a}$.
\ethm
\bprf
Let $q=a-(m+1)\delta$ and $\underline{\Delta}=(n_1,\dots,n_{\tau+1})$ be an $(a,b,\tau,n, r)$-dispersion vector of rate $R_{\max}(a,b,\tau)$ with $\sum\limits_{i=t_0}^{t_0+b-1} n_i=r$. From Lemma~\ref{Lem: sum_r} such a dispersion vector always exists. Let $T=[t_0,t_0+b-1]$ and 
\bea \label{Eq:x_defn}
\resizebox{.45 \textwidth}{!} 
{
$x= \min\left\{\sum\limits_{i=ub+1}^{ub+\epsilon} n_i+\sum\limits_{i=vb+\epsilon+1}^{vb+\delta} n_i ~\Big{|}~  \epsilon \in [1,\delta], u,v \in [0,m], u \le v \right\}.$
}
\eea
Let $u_0$,$v_0$ and $\epsilon_0$ be such that
\bea \label{Eq:part_1}
x=\sum\limits_{i=u_0b+1}^{u_0b+\epsilon_0} n_i+\sum\limits_{i=v_0b+\epsilon_0+1}^{v_0b+\delta} n_i.
\eea 
Since $\underline{\Delta}$ satisfies \eqref{Eq:B}, we have 
\bea
\label{Eq:part_2}
\sum\limits_{i=jb+1}^{(j+1)b} n_i \le r, ~~~\forall j \in [0,u_0-1],\\
\label{Eq:part_3}
\sum\limits_{i=jb+\epsilon_0+1}^{(j+1)b+\epsilon_0} n_i \le r, ~~~\forall j \in [u_0,v_0-1],\\
\label{Eq:part_4}
\sum\limits_{i=jb+\delta+1}^{(j+1)b+\delta} n_i \le r, ~~~\forall j \in [v_0,m-1].
\eea 
Adding \eqref{Eq:part_1}, \eqref{Eq:part_2}, \eqref{Eq:part_3} and \eqref{Eq:part_4} gives an upper bound on $n$,
\bea \label{Eq:n}
n=\sum\limits_{i=1}^{\tau+1} n_i \le mr+x.
\eea 
We will now prove an upper bound on $\frac{x}{r}$ to derive an upper bound on the rate. Let $A_j=[jb+1,jb+\delta]$ for $j \in [0, m]$ and $J = T \cap \left( \cup_{j=0}^m A_j\right)$.
Define $\hat{A}$ to be the set of indices of $q$ largest elements from set $T\setminus J$. It is possible to pick $q$ elements from $T\setminus J$ as $|J| = \delta$ and $|T \setminus J| = b-\delta \ge a-(m+1)\delta = q$. We construct a set $A$ comprising of $a$ elements given by $ A = \cup_{j=0}^m A_j \cup \hat{A}$. Note that the sets $A_j$'s and $\hat{A}$ are all disjoint by definition. This implies that:
\bea
\label{Eq:sum}
\sum\limits_{i \in A} n_i = \sum\limits_{j =0}^m \sum\limits_{i \in A_j} n_i + \sum\limits_{i \in \hat{A}} n_i.
\eea
Let $w \in [0, m-1]$ such that $wb < t_0 \le (w+1)b $, then $J = T \cap (A_w \cup A_{w+1})$. Let us define:
\bea
\label{Eq:Delta}
\sum\limits_{i \in J} n_i = s.
\eea
By the definition of $x$ in \eqref{Eq:x_defn}: 
\bea
\label{Eq:Delta+}
\sum\limits_{i \in \left( A_w \cup A_{w+1} \right) \setminus J} n_i \ge x ~~\text{and}~~ \sum\limits_{i \in A_j}n_i \ge x, ~\forall j \in [0,m].
 \eea 
As we pick $q$ largest element from $T \setminus J$ to get $\hat{A}$ it follows that:
\bea
\label{Eq:hatA} \sum\limits_{i \in \hat{A}} n_i \ge \frac{(r-s)}{b-\delta}q.
\eea
By substituting \eqref{Eq:Delta}, \eqref{Eq:Delta+}, \eqref{Eq:A} in \eqref{Eq:sum} it follows that:
\bea
\label{Eq:rineq} r &\ge& (m-1)x + x + s+\frac{r-s}{b-\delta}q.
\eea
We will now show that $s \ge x$. 
From equation \eqref{Eq:B} we have
\bean
\sum\limits_{i \in T\setminus J} n_i + \sum\limits_{i \in A_w} n_i &\le& r, \\
(r-s) + x &\le& r \text{ from equations \eqref{Eq:Delta}, \eqref{Eq:Delta+}}.
\eean
Therefore $s \ge x$. Substituting this in equation \eqref{Eq:rineq} we get 
 $$\frac{x}{r} \le \frac{b-\delta-q}{(m+1) (b-\delta)-q}
 = \frac{b-a+m\delta}{(m+1)b-a} = \theta.$$
 Now using \eqref{Eq:n}, we have $\frac{n}{r} \le m+\frac{x}{r} \le  m + \theta.$ Hence $R_{\max}(a,b,\tau)=\frac{n-r}{n} \le \frac{m-1+\theta}{m+\theta}$.
\eprf 
\section{Constructions} \label{Sec:constr}
	\vspace{-0.03in}
In this section we present two dispersion vector constructions.   
As explained in Section~\ref{Sec:GSDE}, streaming codes can be constructed by GSDE of appropriate MDS codes guided by these dispersion vectors. 
\begin{constr}Let $\tau+1=mb+\delta$, where $m \in \mathbb{N}$, $0 \le \delta < b$. We construct dispersion vector $(n_1,\dots,n_{\tau+1})$ as:
	\bea \label{Constr:1}
	n_i=
	\begin{cases} 	
		1 ~~~\text{if}~~~ i\mod b \in [1,a],\\
		0 ~~~\text{otherwise}.
	\end{cases}
	\eea
\end{constr}
\bthm
The dispersion vector defined by \eqref{Constr:1} is an $(a,b,\tau,n,r)$ dispersion vector with $n=ma+\min\{a,\delta\}$ and $r=a$. 
\ethm 
\bprf  Clearly,  $\sum\limits_{i = 1}^{\tau+1}n_i = ma+\min\{a,\delta\}$.
Note that $n_i \le 1$ for all $i \in [1,\tau+1]$. Hence for any $A \subseteq [1,\tau+1]$ with $|A|=a$, $\sum\limits_{i \in A}n_i \le a$. It can also be seen that $\sum\limits_{i = j}^{j+b-1}n_i =a$ for all $j \in [1,\tau+2-b]$. 
\eprf 

The rate of above dispersion vector is  $R=\frac{(m-1)a+ \min\{a,\delta\}}{ma+\min\{a,\delta\}}$, which matches with the rate upper bound in Theorem \ref{Thm:upper_bound1} when $\delta =0 $ or $a \le (m+1)\delta$. Hence we have the following corollary.
\bcor 
If $\delta =0 $ or $a \le (m+1)\delta$, then $R_{\max}(a,b,\tau)=R_{SS}(a, b, \tau)=\frac{(m-1)a+ \min\{a,\delta\}}{ma+\min\{a,\delta\}}$.
\ecor 
The GSDE of suitable MDS code with above dispersion vector results in SS codes presented in \cite{NikRamVajKum}. Since all entries of dispersion vector belong to $\{0,1\}$ here, GSDE is same as SDE for this case. For $\{a=3,b=5,\tau=5\}$ example, this dispersion vector construction gives $(1,1,1,0,0,1)$, which is a $(3,5,5,4,3)-$dispersion vector. Now GSDE of a $[4,1]$ MDS code with $(1,1,1,0,0,1)$ leads to  $(a=3,b=5,\tau=5)$ SS code of rate $\frac{1}{4}$ shown in Fig.~\ref{Fig:SDE}. 
Now we present a dispersion vector whose rate is same as the rate upper bound in Theorem \ref{Thm:upper_bound2} for $a>(m+1)\delta>0$ case. 
\begin{constr}
Let $\tau+1=mb+\delta$ with $m \in \mathbb{N}$, $0 \le \delta < b$. Set $t=\frac{lcm(b-a, m)}{b-a}$ and $\gamma = t\left( 1+ \frac{b-a}{m} \right)$. We note that $t$ and $\gamma$ are always non-negative integers. Now we construct dispersion vector $(n_1,\dots,n_{\tau+1})$ as:
\bea
n_i=
\begin{cases} \label{Constr:2}
\gamma ~~~\text{if}~~~ i\mod b =1,\\
t ~~~\text{otherwise}.
\end{cases}
\eea
\end{constr}
\bthm
If $\delta \neq 0$ and $a > (m+1)\delta$, then the dispersion vector defined in \eqref{Constr:2} is an $(a,b,\tau,n, r)$-dispersion vector with $n = t\left(mb+\frac{(m+1)(b-a)}{m}+ \delta \right)$ and  $r=t\left(b+\frac{b-a}{m}\right)$.
\ethm  
\bprf By definition $\sum\limits_{i=1}^{\tau+1}n_i = (m+1)\gamma + (\tau-m)t  =  t\left(mb+\frac{(m+1)(b-a)}{m}+ \delta \right)=n$. 
Pick an $a-$ element subset $\tilde{A} \subseteq [1,\tau+1]$ such that $jb+1 \in \tilde{A}$ for all $j \in [0,m]$. From $b \ge a$, it follows that $\gamma \ge t$. Then it can be easily seen that   
$\sum\limits_{i \in \tilde{A}}n_i \ge \sum\limits_{i \in A}n_i$,
 for all $A \subseteq [1,\tau+1]$ with $|A|=a$.  Now from the definition of $\tilde{A}$ we have
 $\sum\limits_{i \in \tilde{A}}n_i = (m+1) \gamma + (a-m-1)t  =t\left(b+\frac{b-a}{m}\right)=r$.
Thus, \eqref{Eq:A} holds for all $A \subseteq [1,\tau+1]$ with $|A|=a$. Additionally, it holds with equality for some cases. 
For any $j \in [1,\tau+2-b]$, it is easy to see that 
$\sum\limits_{i=j}^{j+b-1}n_i = \gamma + (b-1)t = t \left( b + \frac{b-a}{m}\right)=r$.
Hence \eqref{Eq:B} holds with equality for all the cases. 
\eprf

For $a > (m+1)\delta > 0$, the dispersion vector defined in \eqref{Constr:2} has rate 
\bean
R &=&  \frac{t\left((m-1)(b+\frac{b-a}{m})+\frac{b-a}{m}+\delta\right)}{t\left (m(b+\frac{b-a}{m})+\frac{b-a}{m}+\delta\right)} = 
\frac{m-1+\theta}{m+\theta}, 
\eean 
where $\theta= \frac{b-a+m\delta}{(m+1)b-a}$. This rate matches with the upper bound in Theorem~\ref{Thm:upper_bound2}, resulting in the corollary given below and thereby completing the proof of Theorem~\ref{Thm:main}.
\bcor Suppose $\delta \ne 0$ and $a > (m+1)\delta$. Then  $R_{\max}(a,b,\tau) = \frac{m-1+\theta}{m+\theta}$ where  $\theta= \frac{b-a+m\delta}{(m+1)b-a}$.
\ecor
If $\delta=0$ or $a \le (m+1)\delta$, then GSS code is defined to be same as SS code. For $a>(m+1)\delta>0$, GSS code is obtained through GSDE of appropriate MDS code with dispersion vector defined by \eqref{Constr:2}. Thus, rate of GSS code $R_{GSS}(a,b,\tau)=R_{max}(a,b,\tau)$. Consider $\{a=3,b=5,\tau=5\}$ example. The dispersion vector given by \eqref{Constr:2} for this case is $(3,1,1,1,1,3)$, which is a $(3,5,5,10,7)-$dispersion vector. The resultant $(a=3,b=5,\tau=5)$ GSS code of rate $\frac{3}{10}$ is given in Fig.~\ref{Fig:GSDE}.

If $b > a > (m+1)\delta >0$, then $\frac{b-a+m\delta}{(m+1)b-a} > \frac{\delta}{a}$ and $R_{GSS}(a,b,\tau)=R_{\max}(a,b,\tau)>R_{SS}(a,b,\tau)$. It follows that GSS codes  have rate strictly greater than rate of SS codes if and only if $b > a > (m+1)\delta >0$.
Table~\ref{Tab:rate} presents some parameters for which GSS code gives rate improvement over SS code. The table also shows field size requirement for those parameters. We note that $n \le \frac{\tau^2}{b}+(\tau+1)b$ for GSS codes and hence a finite field of size $\ge \frac{\tau^2}{b}+(\tau+1)b$ is sufficient in the worst-case.
\begin{table}
	\scriptsize {
		\begin{center}
				\caption{$R_{opt}$ is the optimal rate of an $\{a,b,\tau\}$ code. The field size $q_{opt}$ corresponds to the rate-optimal code construction in \cite{NikDeepPVK}.  $R_{\text{\tiny SS}}, \ q_{\text{\tiny SS}}, \ R_{\text{\tiny GSS}}, \ q_{\text{\tiny GSS}}$ are the respective rates and field size associated to an SS code and GSS code.  }
			\begin{tabular}{ || c || c | c || c | c || c | c || }
				\hline \hline 
				
				$(a,b,\tau)$ & $R_{opt}$ & $q_{opt}$  & $R_{\text{\tiny SS}}$  & $q_{\text{\tiny SS}}$ & $R_{\text{\tiny GSS}}$  & $q_{\text{\tiny GSS}}$ \\ \hline \hline 
				(3,5,5) & 0.375 & 25 &  0.25 & 3  &  0.3 & 9 
				\\ \hline  
				(4,5,10) & 0.583 & 100 &  0.556 & 8  &  0.56 & 24  \\ \hline 
				(5,8,16) &     0.6  & 256 &   0.545 & 10 &    0.558 & 42 
				\\ \hline 
				(9,15,15) &    0.318  & 225  &  0.1 & 9  &    0.25 & 27 \\ \hline 
				(10,18,20) &  0.379 & 400  &  0.231 & 12  &   0.297 & 36  \\ \hline \hline 
			\end{tabular} 
			\label{Tab:rate}
	\end{center}}
	\vspace{-0.3in}
\end{table}
	\vspace{-0.1in}
\section{Conclusion}
Simple streaming codes are generalized here by permitting each packet to contain more than one code symbol from the underlying scalar code.  This results in an increase in code rate that is quantified. The generalized streaming codes presented in this paper use MDS code as the underlying scalar code along with conventional block decoding and hence are simpler to implement than rate-optimal streaming codes known in literature.

%
%
%
\bibliographystyle{IEEEtran}
\bibliography{streaming}
\appendix

\subsection{Proof of Theorem~\ref{Thm:upper_bound1}} 
\label{proof_upper_bound}
Let $\underline{\Delta}=(n_1,\dots,n_{\tau+1})$ be an $(a,b,\tau,n, r)$-dispersion vector  with rate $R_{\max}(a,b,\tau)$. Suppose $\delta < a \le (m+1)\delta$.
Let  $a=p\delta+q$ where $p \in \mathbb{N}, 0 \le q < \delta$. Let us set
\bea
\label{Eq:x}
x= \min\left\{\sum\limits_{i=jb+1}^{jb+\delta} n_i ~\Big{|}~ j \in [0,m]\right\}.
\eea 
We first obtain an upper bound on $n$ in terms of $r$ and $x$. Let $j_{\tiny 0} \in [0,m]$, such that 
\bea \label{Eq:1}
x=\sum\limits_{i=j_{\tiny 0}b+1}^{j_{\tiny 0}b+\delta} n_i.
\eea 
Since $\underline{\Delta}$ satisfies \eqref{Eq:B},
\bea 
\label{Eq:2}
\sum\limits_{i=jb+1}^{(j+1)b} n_i \le r, ~~~\forall j \in [0,j_{\tiny 0}-1],\\
\label{Eq:3} 
\sum\limits_{i=jb+\delta+1}^{(j+1)b+\delta} n_i \le r ~~~\forall j \in [j_{\tiny 0},m-1].
\eea
Adding \eqref{Eq:1}, \eqref{Eq:2} and \eqref{Eq:3},
\bea \label{Eq:4} 
n=\sum\limits_{i=1}^{\tau+1} n_i \le mr+x.
\eea 
Now we derive an upper bound on $\frac{x}{r}$ and then use it to obtain the rate upper bound. Define $A_j=[jb+1,jb+\delta]$ for all $j \in [0,m]$.  From \eqref{Eq:x} we get
\bea \label{Eq:5}
\sum\limits_{i \in A_j}n_i \ge x, ~~~\forall j \in [0,m].
\eea  

Let $\hat{A}$ be the set of indices of $q$ largest elements from $\{n_i ~\mid~ i \in [1,\delta]\}$.  Then by pigeonhole principle,
\bea \label{Eq:6}
\sum\limits_{i \in \hat{A} }n_i \ge \frac{qx}{\delta}.
\eea 
Let us define $A= \begin{cases} \bigcup\limits_{j=0}^{i=m}A_j ~~~\text{if}~ a=(m+1)\delta,\\
\bigcup\limits_{j=1}^{i=p}A_j \bigcup \hat{A} ~~~\text{otherwise}.
\end{cases}$

Note that $|A|=a$ in both cases since $A_{j_1} \cap A_{j_2} =\phi$ for any two distinct $j_1, j_2 \in [0,m]$ and $\hat{A} \cap A_{j} =\phi$ for all $j\in [1,m]$. If $a=(m+1)\delta$, it follows from \eqref{Eq:A} and \eqref{Eq:5}  that 
\bean 
r \ge \sum\limits_{i \in A}n_i \ge (m+1)x \implies \frac{x}{r} \le \frac{1}{(m+1)}=\frac{\delta}{a}. 
\eean 
Now for $\delta < a < (m+1)\delta$, using \eqref{Eq:2}, \eqref{Eq:5} and \eqref{Eq:6}, we obtain 
\bean 
r \ge \sum\limits_{i \in A}n_i \ge px+ \frac{qx}{\delta}
\implies \frac{x}{r} \le \frac{\delta}{p\delta+q}=\frac{\delta}{a}.
\eean 
From \eqref{Eq:4} we get 
\bean 
\frac{n}{r} \le m +\frac{x}{r} \le m+\frac{\delta}{a}.
\eean 
Hence when $\delta < a \le (m+1)\delta$, 
\bean 
R_{\max}(a,b,\tau)=\frac{n-r}{n} \le \frac{m-1+\frac{\delta}{a}}{m+\frac{\delta}{a}} = R_{SS}(a,b,\tau).
\eean 

Now consider the $\delta =0$ case. Since $\underline{\Delta}$ satisfies \eqref{Eq:B}, $\sum\limits_{i=jb+1}^{(j+1)b}n_i \le r, \forall j \in [0,m-1]$. Here $\tau+1=mb$ and hence we have $$n=\sum\limits_{i=1}^{\tau+1}n_i \le mr.$$
Therefore, 
\bean 
R_{\max}(a,b,\tau) \le \frac{m-1}{m}=R_{SS}(a,b,\tau)
\eean 
 when $\delta=0$.

\subsection{Proof of Lemma~\ref{Lem: sum_r}}
\label{proof_sum_r}
The Lemma~\ref{Lem: sum_r} follows from repeated application of Lemma~\ref{Lem: sum_increase}.

\blem \label{Lem: sum_increase} Let $\underline{\nu}=(n_1,\dots,n_{\tau+1})$ be an $(a,b,\tau,n, r)-$dispersion vector with rate $R_{max}(a,b,\tau)$. Define $s = \max\left\{ \sum\limits_{i=j}^{j+b-1} n_i ~\Big{|}~ j \in [1, \tau+2-b]\right\}$ and suppose $s < r$ . Then there exists an $(a,b,\tau,n,r)-$dispersion vector  $\underline{\mu}=(m_1,\dots,m_{\tau+1})$ with $\sum\limits_{i=j}^{j+b-1} m_i = s+1$ for some $j \in [1,\tau+2-b]$.
\elem  
\bprf 
It is always possible to pick a subset $I \subseteq [1,\tau+1]$ with $|I|=a$ such that $n_i \ge \max\{n_j ~\mid~ j \in [1,\tau+1]\setminus I \}$ for all $i \in I$. Now suppose $\sum\limits_{i \in I}n_i < r$. Then, it follows that $\sum\limits_{i \in A}n_i < r$, for all $A \subseteq [1,\tau+1]$ with $|A|=a$. Since $s < r$, $(n_1,\dots,n_{\tau+1})$ doesn't satisfy \eqref{Eq:B} with equality for any $j \in [1,\tau+2-b]$. This contradicts the fact that  $(n_1,\dots,n_{\tau+1})$ is an $\{a,b,\tau,n,r\}-$dispersion vector. Hence, 
\bea \label{Eq:sum_max_A}
\sum\limits_{i \in I}n_i = r.
\eea 
Define $z=\min\{n_i ~\mid~ i \in I\}$ and  $J=\{i \in I ~|~ n_i>z\}$. Now we will argue that $n_i=z$ for all $i \in [1,\tau+1]\setminus J$. Let $p \in [1,\tau+1]\setminus J$ with $n_p <z$. Assign $n'_p=n_p+1$ and $n'_i=n_i, \forall i \in [1,\tau+1]\setminus \{p\}$. For any $j \in [1,\tau+2-b]$, 
\bean 
\sum\limits_{i=j}^{j+b-1} n'_i \le \sum\limits_{i=j}^{j+b-1} n_i+1 \le r.
\eean  
Since $n'_p \le z$, 
\bean 
\sum\limits_{i \in A}n'_i \le \sum\limits_{i \in I}n_i = r
\eean 
for all $A \subseteq [1,\tau+1]$ with $|A|=a$. We also have
\bean 
\sum\limits_{i=1}^{\tau+1} n'_i=\sum\limits_{i=1}^{\tau+1} n_i+1=n+1.
\eean 
Therefore $(n'_1,\dots,n'_{\tau+1})$ is an $(a,b,\tau,n+1, r)-$dispersion vector. This dispersion vector has rate $\frac{n+1-r}{n+1}>\frac{n-r}{n}=R_{max}(a,b,\tau)$, which is a contradiction. Hence, 
\bea \label{Eq:z}
n_i=z, ~~~\forall i \in [1,\tau+1]\setminus J.
\eea 
Suppose $J=\phi$, resulting in $n_i=z$ for all $i \in [1,\tau+1]$. Then, from \eqref{Eq:sum_max_A} we have $r=az$. It follows from statement of Lemma that $s=bz<r=az$. Since $z \ge 0$ and $b \ge a$, this results in a contradiction. Therefore $J$ is non-empty. We note that $1 \le |J| \le |I|-1 <a$ and $n_i =z,~\forall i \in I \setminus J$. From \eqref{Eq:sum_max_A}, we get   
\bea \label{Eq:J}
\sum\limits_{i \in J} n_i+(a-|J|)z = r.
\eea 
Let $\sum\limits_{i=t}^{t+b-1} n_i=s$ and $T=[t,t+b-1]$. Since $|T|=b$ and $|J|<a$, it follows from \eqref{Eq:z} that there exists some $q \in T \setminus J$ with $n_q=z$. If $J \subseteq T$, then by \eqref{Eq:z}  and \eqref{Eq:J} we have 
\bean 
\sum\limits_{i=t}^{t+b-1} n_i=\sum\limits_{i \in J} n_i+(b-|J|)z \ge r>s,
\eean 
which is not possible. Hence, there exists some $p \in J \setminus T$ such that $n_p >z$. 
With this notation we define the new dispersion vector $\underline{\mu}=(m_1, \dots, m_{\tau+1})$ as:
\bean
m_p = n_p - 1,~~~ m_q = n_q + 1 ~~~\text{and}
\eean 
\bean 
m_i = n_i, ~~~\forall i \in [1,\tau+1]\setminus \{p,q\}.
\eean 
\subsubsection*{Claim 1} $\underline{\mu}$ is an $(a,b,\tau,n,r)-$dispersion vector with rate $R_{max}(a,b,\tau)$. \\
\subsubsection*{Proof of Claim 1} 
Since $m_p=n_p-1\ge z \ge 0$,  $\underline{\mu}$ is a $(\tau+1)$-tuple of non-negative integers. Now from definition of $\underline{\mu}$ it follows that 
\bean 
\sum\limits_{i=1}^{\tau+1} m_i = \sum\limits_{i=1}^{\tau+1} n_i=n,
\eean 
thus satisfying \eqref{Eq:total_sum}. For all $j \in [1,\tau+2-b]$,
\bean 
\sum\limits_{i=j}^{j+b-1} m_i \le \sum\limits_{i=j}^{j+b-1} n_i+1 \le s+1 \le r.
\eean 
Thus $\underline{\mu}$ satisfies \eqref{Eq:B}. 

If $\sum\limits_{i \in A} n_i=r$ for $A \subseteq [1,\tau+1]$ with $|A|=a$, then $J \subseteq A$. Otherwise,~\eqref{Eq:A} is violated for $\underline{\nu}=(n_1,\dots,n_{\tau+1})$. It can be easily seen that 
\bean 
\sum\limits_{i \in J} m_i = \sum\limits_{i \in J} n_i - 1
\eean
\bean 
~~~\text{and}~~~  \sum\limits_{i \in B} m_i \le \sum\limits_{i \in B} n_i + 1, ~\forall B \subset [1,\tau+1]. 
\eean 
For all $A \subseteq [1,\tau+1]$ such that $|A|=a$ and $J \subseteq A$,  
\bean 
\sum\limits_{i \in A} m_i &=&  \sum\limits_{i \in J} m_i + \sum\limits_{i \in A\setminus J} m_i \\ &\le& \sum\limits_{i \in J} n_i - 1 + \sum\limits_{i \in A\setminus J} n_i + 1 \\ &\le& \sum\limits_{i \in J} n_i - 1 + (a-|J|)z+1=r.
\eean  
For $A \subseteq [1,\tau+1]$ such that $|A|=a$ and $J \subsetneq A$, $\sum\limits_{i \in A} n_i < r$. This means 
\bean 
\sum\limits_{i \in A} m_i \le \sum\limits_{i \in A} n_i + 1 \le r,
\eean 
for all $A \subseteq [1,\tau+1]$ with $|A|=a$ and $J \subsetneq A$. Thus, \eqref{Eq:A} holds for $\underline{\mu}$. 

It is possible to pick an $A \subset [1,\tau+1]$ with $|A|=a$ and $J \cup \{q\} \subseteq A$, then we have  
\bean 
\sum\limits_{i \in A} m_i &=& \sum\limits_{i \in J} m_i+ m_q+  \sum\limits_{i \in A\setminus (J \cup \{q\})} m_i \\ &=& \sum\limits_{i \in J} n_i-1+z+1+(a-|J|-1)z = r.
\eean 
Thus, $\underline{\mu}$ is an $(a,b,\tau,n, r)-$dispersion vector. This completes the proof of Claim 1. 

Since $q \in T$ and $p \notin T$, it can be easily seen that $\sum\limits_{i=t}^{t+b-1} m_i=\sum\limits_{i=t}^{t+b-1} n_i +1=s+1$.
\eprf

\end{document}